\documentstyle[12pt,aaspp4,psfig]{article}
\def\lsim{~\rlap{$<$}{\lower 1.0ex\hbox{$\sim$}}}
\def\gsim{~\rlap{$>$}{\lower 1.0ex\hbox{$\sim$}}}
\def\p3m{P$^3$M}

\begin{document}

\title{DARK MATTER, DISCRETENESS AND COLLISION ERROR IN COSMOLOGICAL N-BODY
SIMULATIONS}
\author{Randall J. Splinter\altaffilmark{1},
Adrian L. Melott\altaffilmark{2},  and Sergei F. Shandarin\altaffilmark{2}}.

\affil{randal@convex.hp.com, melott@kusmos.phsx.ukans.edu,
  sergei@kusmos.phsx.ukans.edu}

\begin{abstract}

We report on a series of tests of agreement between three types of N-body
simulations: PM, P$^3$M, and Tree codes. We find good agreement in both the
individual and the statistical properties only on scales larger than the
mean interparticle separation.  As a result, we question most numerical results
at and below below galaxy scales, either concerning primordial dark matter
or baryonic matter coupled to it by gravitation.

\altaffiltext{1}{Current Address: Hewlett--Packard Company, High
Performance Computing Division, 20 Perimeter Summit Blvd, MS 1904, Atlanta,
GA 30319--1417}

\altaffiltext{2}{Department  of Physics  and Astronomy, University  of
Kansas, Lawrence, KS 66045}

\end{abstract}


\section{Introduction}

If the dark matter in the Universe is some sort of weakly interacting particle,
or even a population of primordial black holes, then artificial
discreteness effects  of all kinds should be throughly suppressed in numerical
simulations of its gravitational clustering. Their contribution both to
errors in the initial conditions and two-body scattering in the
dynamical evolution are negligible in the Universe and therefore should
be negligible in the simulation.
If not, the results (including
those of baryons, entrained in the gravitational field) cannot be relied upon
to be  consequences of physical initial conditions -- regardless of whether
they resemble
astronomical observations.

It has been argued, based on a variety of numerical studies, that there are
real problems with the HFLMR (High Force Low Mass Resolution) approach which
dominates the cosmological N--body methodology. (Peebles, et al. 1989; Melott
1990; Suisalu and Saar 1995; Kuhlman et al. 1996; Melott et al. 1997; Park
1997). It is clear from
these studies that there are unphysical scattering processes taking place in
HFLMR codes, but the precise effects are
not clear. Recently Craig (1997), Klypin et al (1997), and Moore et al. (1997)
have argued
convincingly that the central regions of dark matter halos have been seriously
misrepresented by numerical results due to lack of resolution.
We present here some preliminary results of a much larger study (Splinter et
al.
1998) which explores the limitations of mass resolution in cosmological
clustering simulations.

We use PM (Hockney \& Eastwood 1988; Melott 1981, 1986), AP$^3$M
(Couchman 1991), and Tree (Suginohara et al. 1991; Suto 1993) codes.
The \p3m code had adaptive smoothing turned off since we wish to
compare a standard \p3m method.  The Tree runs use the fixed smoothing
length in comoving coordinates, and we set a tolerance parameter
$\theta=0.2$ which is considerably smaller (and thus more accurate)
than conventional choices ($\theta=0.5\sim0.75$). $\theta$ merely controls how
far the tree expansion is carried, and thus the accuracy of long-range forces.

The initial power spectrum in all cases was $P(k) \propto k^{-1}$ up
to some cutoff, in most cases at the Nyquist frequency $k = 16k_f$
dictated by the runs with the fewest particles.  Realization of the
corresponding density field was generated using the Zel'dovich
approximation (Zel'dovich 1970) to perturb the particles from their
initial lattice (Doroshkevich et al. 1980).  All the models are
evolved in the Einstein -- de Sitter universe ($\Omega_0=1$).  The
comparisons were performed at three different epochs when the
nonlinear wavenumber $k_{\rm nl}$ becomes $16k_{\rm f}$, $8k_{\rm f}$,
and $4k_{\rm f}$, where $k_{\rm nl}=k_{\rm nl}(A)$ is defined as
\begin{equation}
\sigma^2(k_{\rm nl},A) = A^2 \int^{k_{\rm nl}}_0 P(k) d^3k = 1 .
\end{equation}
In the above $A$ denotes the expansion factor (unity at the initial
condition), and these values of $k_{\rm nl}$ correspond to the epochs
$A=22.36$, $42.13$, and $92.20$, respectively.  The latest moment we
studied corresponds to nonlinearity on the largest scale which does
not suffer from finite-volume boundary condition problems (Ryden \&
Gramman 1991; Kauffmann \& Melott 1992).  The specific runs presented here
and the model parameters are shown in Table 1.  We note that PM codes,
which have been extensively used in most physical applications with
large numbers of particles, are much faster than the other two types.
Thus our 128$^3$ PM runs took much less CPU time than even the 32$^3$
Tree or \p3m runs.  The typical limitation on PM runs is memory or
disk space, while CPU time is the typical limitation on \p3m or Tree runs.
$a$ is the absolute scale of force resolution.
We define $\epsilon = a {\overline
n_{\rm sim}^{1/3}}$ so that $\epsilon = 1$ corresponds to
smoothing at the mean interparticle separation.
HFLMR codes usually run with $\epsilon < 1$, but we will
examine their behavior over a range in N and $\epsilon$, pushing them
toward $\epsilon = 1$ by increasing the number of particles while
keeping the absolute scale of force resolution $a$ constant.  So far
as we know, this crucial experiment has not previously been done with
HFLMR codes.

Our primary strategy is therefore to highlight the largely unexplored
mass resolution issue by varying the number of particles while keeping
$a$ constant (it is not {\it exactly} constant because in fact the
shape of the short-range softening function is different in all three
codes). Within a given code $a$ will be constant, so we can spot
trends.  Between codes softening will be of comparable size.  We can define
$r_{50}$ as the radius where the force drops to 50\% of the Newtonian value.
For the $P^3M$ code, $r_{50}=0.92a$. For the Tree code $r_{50}=0.87a$. For our
PM code, $r_{50}=0.95$ grid unit, albeit with considerable scatter in the
softened zone.

In most
cases we set $a = 1$ (in units where the box size is 128) in both the
Tree and \p3m codes, and run PM with a 128$^3$ mesh.  (The $P^3M$ runs had one
mesh cell per particle, but this is not the factor determining force
resolution there.) It is typical of
HFLMR codes to have $a$ considerably less than $\bar{l}_{sep}$ , the
mean interparticle separation over the entire simulation box.  In most
uses of PM codes. the opposite is true (although in gravitational applications
it has become customary to have $\epsilon$ as small as $1$ or $0.5$).

It is important to note that one cannot represent initial power to
higher than the Nyquist wavenumber of the particles or the mesh of the
FFT used to impose the initial conditions, whichever is worse (in fact
the latter is rarely the problem in cosmology).  Therefore most of our
runs only have initial power up to $k_{\rm c} = 16k_{\rm f}$, so that
comparisons of only the effects of divergent numerical integration can
be done. We show here results from a full $128^3$ PM simulation
with this power spectrum, and
$32^3$ simulations with the same {\it force}
resolution, different mass resolution, and the same initial conditions.
In order to explore the effects of having initial
power at higher wave-numbers which is only possible with more
particles, we have one PM run with $k_{\rm c} = 64 k_{\rm f}$.

We now have for the first time a large number of runs using 3
different codes with identical initial conditions, identical force
resolution, but varying mass resolution.  In particular the number of
particles varies widely enough to study the same physical system with
values of $\epsilon$ typical of HFLMR codes as well as to study the
same system with a PM code with similar force resolution {\it and}
matching mass resolution ($\epsilon = 1$).

It is important to note time-step limitations. Elementary principles of
numerical
stability require that no particle move more than a fraction of a softening
length $a$, or
about one--half mesh unit for $PM$, in a single time-step. This condition was
amply
enforced for all three codes.

\section{Results}

Figure 1 shows power spectra computed on the 128$^3$ mesh appropriate to the
force resolution, normalized such that a Poisson distribution with 128$^3$
particles has $P(k)\sim 1$. The dotted line shows the spectrum of the $32^3$
initial conditions. It looks unusual because such spectra are not usually shown
out to the force resolution scale -- even though later results usually are
shown
to this scale. Other types of unperturbed particle arrangements than our
lattice
will have other features, but the overall amplitude is constrained to be
similar.
A full complement of 128$^3$ particles would allow the power law to continue
across the plot.

It should be noted that in all HFLMR codes there are features due to particle
discreteness present beyond the particle Nyquist frequency, which are resolved
and evolved. Whatever the particle arrangement, they are not random phase
and cannot evolve as the initial conditions do since they are tied to the
particles.

Above this we see spectra for three evolved stages.
The first (corresponding to the onset of nonlinearity at the mass resolution
scale of the 32$^3$ particle simulations) and the last (without serious
boundary condition problems) of these are shown on the left side as
ratios to the power in one of them, in order to clarify differences.  It
can be seen that in the first of these
that the spectra agree well at low $k$ (linear theory) and very
high $k$
(pure mode coupling) frequencies, but disagree between. In the range of
$10<k<20$ (in units of the fundamental $k_f$),
there are three classes of spectra: those with 32$^3$ particles (bottom
group), those with the same initial conditions but more particles (middle
group) and the one which had full initial power down to $k=64$ (top group).
There is nearly a factor of two range here. (In this and following measures we
compute spectra from an initially identical $32^3$ subset of the 128$^3$
run to suppress sampling
differences). By the final stage (as evolved as is safe with periodic boundary
conditions) the differences have moved to high $k$. If only force resolution
mattered, these would be identical to $k=64$. In fact, they diverge around
$k=20$, close to the limit set by mass resolution. We cannot say which run is
correct, and we did not even find a monotonic trend with mass resolution.

An arbitrary density field is described by both the amplitudes and phases of
its
Fourier coefficients. We now examine the phase agreement between various codes.
Figure 2 examines $<cos\; \theta>$, where $\theta$ is the difference in phase
angle between the same Fourier coefficient in different simulations and the
average is over all wave-numbers with the same amplitude $|k|$.
Simulations which agree have $<cos\; \theta>=1$; totally uncorrelated results
have $<cos\; \theta>=0$. The open circles show the
$128^3$ PM and P$^3$M runs with identical initial conditions compared.
(We do not have a $128^3$ Tree run but results with $64^3$ and
$32^3$ runs shown in Splinter et al. 1998 support the same trend.) The bottom
group (filled squares) shows the $128^3$ particle PM run with full initial
conditions down to $k=64$ compared with all other runs. In order to reduce
confusion, we do not show our other results, which lie in between. Thus
(a) As the number of particles are increased, with force
resolution and initial conditions held constant, different codes agree
better. (b) Even when they agree (given identical initial conditions),
none of them agree strongly on small scales with a run that had the full range
of proper initial conditions as allowed by having more particles.

In Table 2, we show the density correlation coefficient
$k_{12}={<\delta_1\delta_2>\over (\sigma_1\sigma_2)^{1/2}}$ where $\delta$ are
the density contrasts in two simulations and $\sigma$ their RMS values. Above
the
diagonal we
show values on the $128^3$ mesh; below the $32^3$ mesh. Below the
diagonal, results approach $k=1$, indicating near perfect agreement. Above
the diagonal, again recalling that the $k_c=16$ PM run, the Tree, and P$^3M$
runs have the same initial conditions, we see that these runs agree best when
the $128^3$ runs are compared. Lastly, the $k_c=64$ PM run does not agree
strongly with any of the others. Again we conclude that integration errors are
greatly reduced when more particles are included, but that the inclusion of
perturbations on small scales is also important.
The precise effect of ignoring them is spectrum-dependent.

\section{Does It Matter?}

The reader is referred to the center and
bottom center images of Figure 7 in Beacom et al. (1991) for another
example of the importance of initial power on small scales even when
it is deep in the nonlinear regime.  There, as well as in Melott \& Shandarin
(1993), we emphasized the effectiveness of the transfer of power from long to
short waves. The general position and orientation of objects is
determined by initial perturbations on that comoving scale and larger, so
smaller perturbations are ignorable for this purpose. See also Little et al.
(1991), Evrard and Crone (1992). However, here we stress that the {\it internal
structure} of these objects will vary depending on smaller--scale
perturbations. If we wish to study that internal structure, the smaller--scale
initial perturbations  must be
present and properly evolved.

We have shown that even with nearly identical force resolution, different
N--body codes differ in their results below the mean interparticle separation.
They converge either by smoothing on this larger scale or by
putting in more particles so that the scales are the same. Codes which give
different
answers cannot all be correct. On the other hand, it might be that the
results on small scales are statistically equivalent, in the sense
that quantities of interest computed from the ensembles are the same.
This turns out not to be true. Even the autocorrelation function is affected by
mass resolution; phase sensitive measures are more strongly affected. We
present here
only one simple set of measures, based on the mass density.

The percolation code of Yess and Shandarin (1996) is used to identify connected
regions on the $128^3$ mesh above a given density threshold. In Figure 3 we
show the ratio of total mass in regions above a given density contrast to the
value found in the $k_c=64$ PM simulation at the first and last of our evolved
stages. The situation is time--dependent and
complex to describe. Clearly there are major differences, but the codes agree
in low density regions. The total mass at the high density limit varies by an
order of magnitude. At the earlier stage all curves were below
unity in the region of $\delta \sim 50$, suggesting that a generation of
structures were missed by the absence of correct initial conditions found only
in one run. Later
we find that within a code type, a higher particle density results in
higher peak densities. Clearly, high density regions are not trustworthy! We
found that with binning on the $32^3$ mesh (not shown) the curves agree well;
the maximum difference there is about 25\%.

We have examined infall velocities and velocity dispersions (Splinter et al.
1998) and found measurable differences. All the differences we find are
systematic errors, so no error bars are shown. The differences should be
measured against the desired accuracy one wants to get from the computations.

We cannot easily compare our results with others, because no study
has included our variety of codes, mass resolution, and inclusion of
normally absent small--scale perturbations.  The most similar work is
the unpublished study coordinated by D.H. Weinberg.  An analysis similar to
our velocity studies (not shown) produced very similar results.  Efstathiou
et al. (1988) studied evolution of power--law initial conditions in
a P$^3$M code, using primarily low--order or averaged statistics.
They did find fluctuations of order 50\% in the rescaled multiplicity
function (their Figure 9); while not equivalent, this is roughly
compatible with the nature of our results in Figure 3.

We believe the safest course to follow is to restrict attention to scales above
the mean particle separation. Given current computer technology and volumes
large enough to respect boundary conditions, this implies scales of order 100
to 200 kpc. With nested--grid codes (eg. Splinter 1996) or other schemes to
approximate an external zone,
this may be considerably improved. However, we caution that careful testing
is needed in the nonlinear regime. It is not enough to produce something that
resembles our Universe; we must have confidence that it is a consequence of
the initial conditions that were supposed to be modeled.

At the present time, we are quite skeptical of nearly all numerical results on
early galaxy
formation and the inner parts of dark matter halos,
as they are typically below the mean interparticle separation of the
simulations in question.

\bigskip

\acknowledgements

ALM and SFS acknowledge financial support from the NSF--EPSCoR program,
NASA grant NAG5-4039, and
computing resources at the National Center for Supercomputing Applications.
We thank Yasushi Suto for Treecode simulations permission to discuss common
results.

\newpage

{\tiny
%
\begin{deluxetable}{ccccc}
\tablenum{1}
\tablecaption{Model Parameters for the Test Cases}
\tablehead{
\colhead{Code}&
\colhead{}&
\colhead{N}&
\colhead{$\bar{l}_{sep}$\tablenotemark{a}}&
\colhead{$\epsilon_{force}$\tablenotemark{b}}}
\startdata
PM      & $k_c=16$ & $128^3$ & 1.0 & 1.0    \nl
        & $k_c=64$ & $128^3$ & 1.0 & 1.0    \nl
P$^3$M  & $k_c=16$ & $32^3$  & 4.0 & 0.25   \nl
        & $k_c=16$ & $128^3$ & 1.0 & 1.0    \nl
Tree    & $k_c=16$ & $32^3$  & 4.0 & 0.25   \nl
\tablenotetext{a}{$\bar{l}_{sep}$ is the  mean particle separation  in
grid  cell  units.}
\tablenotetext{b}{$\epsilon_{force}$ is  in units of the mean particle
separation, $\bar{l}_{sep}$.   Note that  for  all runs here
$a \equiv  \epsilon_{force} \times \bar{l}_{sep} = 1$.}
\enddata
\end{deluxetable}

\begin{deluxetable}{lcccccccc}
\tablenum{2}
\tablecolumns{9}
\tablewidth{0pt}
\tablecaption{Cross--Correlations at Final Stage}
\tablehead{
\colhead{}&
\multicolumn{2}{c}{PM}&
\colhead{} &
\multicolumn{2}{c}{P$^3$M}&
\colhead{} &
\multicolumn{2}{c}{Tree}\\
\cline{2-3} \cline{5-6} \cline{8-9}\\
\colhead{}&
\colhead{$k_c=16$} &
\colhead{$k_c=64$} &
\colhead{}&
\colhead{$128^3 \epsilon=1.0$}&
\colhead{$\epsilon=0.25$} &
\colhead{}&
\colhead{$\epsilon=0.25$}}
\startdata
PM ($k_c=16$)                 & ---  & 0.65 & & 0.83  & 0.64 & & 0.52 & \nl
PM ($k_c=64$)                 & 0.96 & ---  & & 0.63  & 0.57 & & 0.47 & \nl
P$^3$M $128^3 (\epsilon=1.0$) & 0.99 & 0.95 & & ---   & 0.66 & & 0.50 & \nl
P$^3$M($\epsilon=0.25$)       & 0.96 & 0.94 & & 0.97  & ---  & & 0.70 & \nl
Tree($\epsilon=0.25$)         & 0.92 & 0.90 & & 0.92  & 0.96 & &  --- & \nl
\enddata
\end{deluxetable}

\clearpage
\begin{figure}
\begin{center}
  \leavevmode\psfig{figure=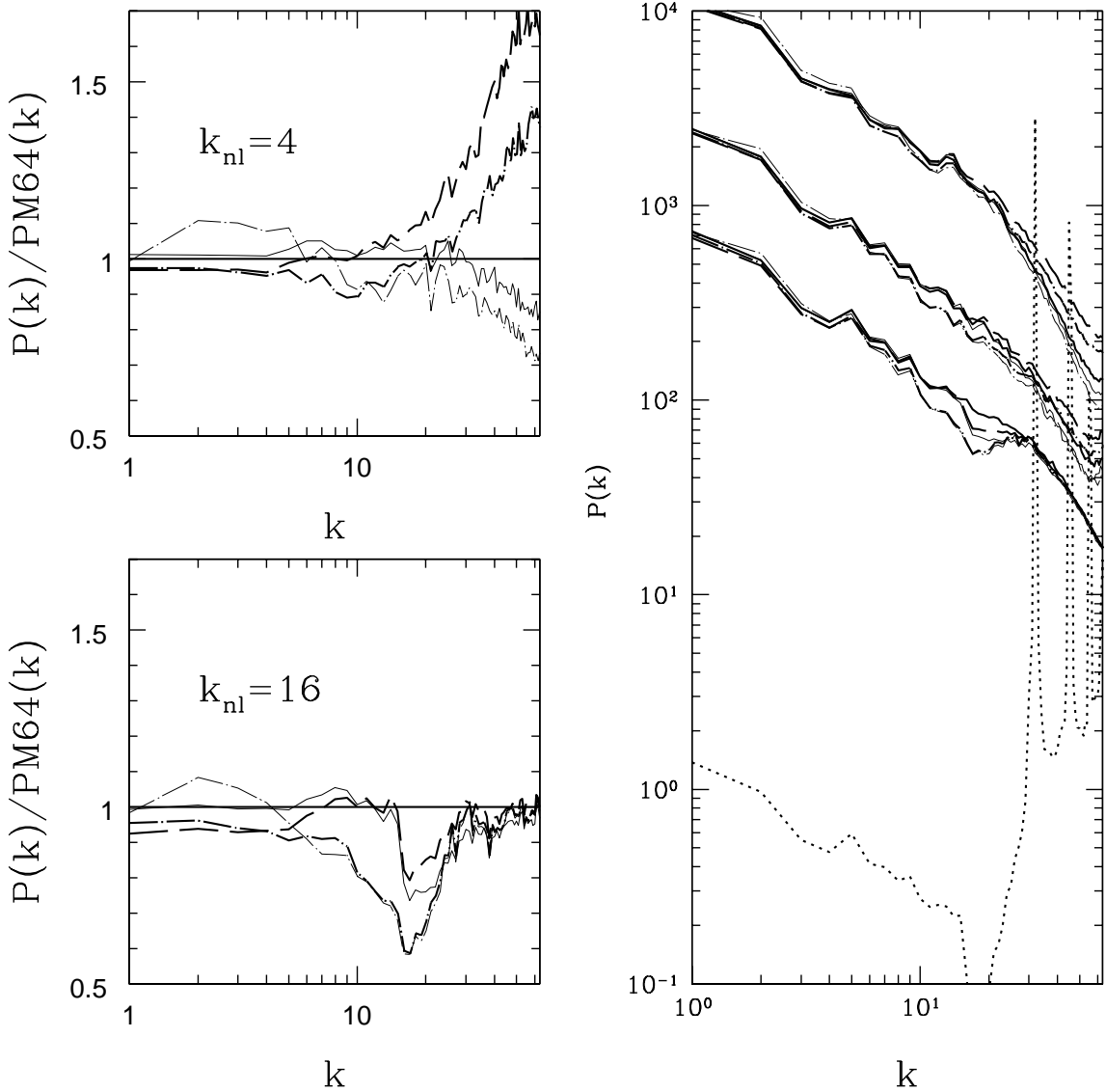,height=16cm}
\end{center}
\caption{
  Right panel:
  The power spectrum of our initial conditions and three successive
  evolved stages
  for all our simulations, evaluated from 32$^3$ particles on a
  128$^3$ mesh (their force resolution).  The normalization is such that
  a Poisson distribution of 128$^3$ particles would converge to $P =
  1$.  The light solid line is the $k_{\rm c} = 16k_{\rm f}$ PM run;
  the heavy solid line is the $k_{\rm c} = 64k_{\rm f}$ PM run.  Other
  heavy lines are \p3m runs and light lines are tree code runs.
 The dotted lines are the initial conditions  for the
 32$^3$ particle runs; the spikes are discreteness
 effects due to the finite number of particles.
 Other lines: Longdash is the \p3m run with 128$^3$ particles,
 and the longdash-dot lines are the \p3m and Tree runs with
 32$^3$ particles.
 Left panel: The ratio of the power in a given model to that in our
 fiducial $k_{\rm c} = 64k_{\rm f}$ PM run at the first (bottom)
 and last (top) evolved stage.
\label{fig:fig1}}
\end{figure}

\begin{figure}
\begin{center}
  \leavevmode\psfig{figure=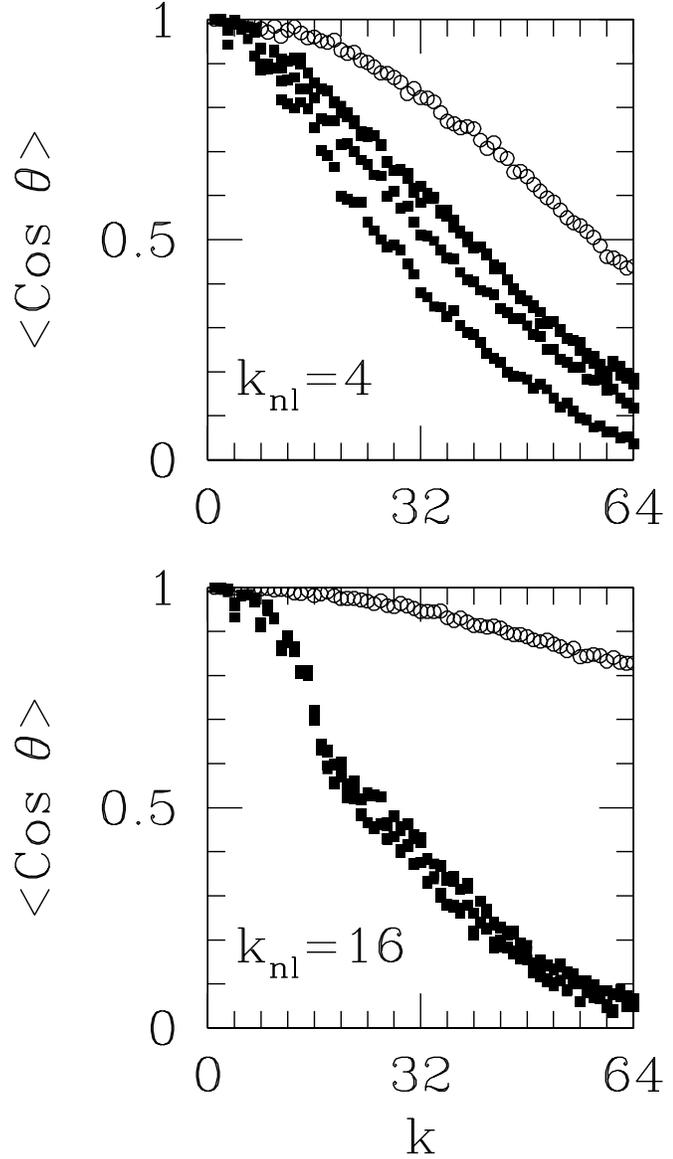,height=16cm}
\end{center}
\caption{
  The successive plots contain all the data for the averaged phase
  agreement between {\it all} of the simulations runs at the same
  stage; $< cos\; \theta > $ is defined in the text and is $1$ for
  agreement and $0$ for uncorrelated phases.
  The filled squares indicate cross-correlation of other runs
  against the PM run which
  continued the power-law perturbations to wave-numbers impossible for
  small numbers of particles ($k_{\rm c} = 64k_{\rm f}$).
  Open circles represent cross-correlation between the 128$^3$ P$^3$M
  run and the PM run with $k_{\rm c} = 16k_{\rm f}$ (also 128$^3$ particles).
  All other possible cross-correlations here (not shown) lie between these
  extremes.
  As before, the top panel is the last evolved stage.
\label{fig:fig2}}
\end{figure}
\begin {figure}
\begin{center}
  \leavevmode\psfig{figure=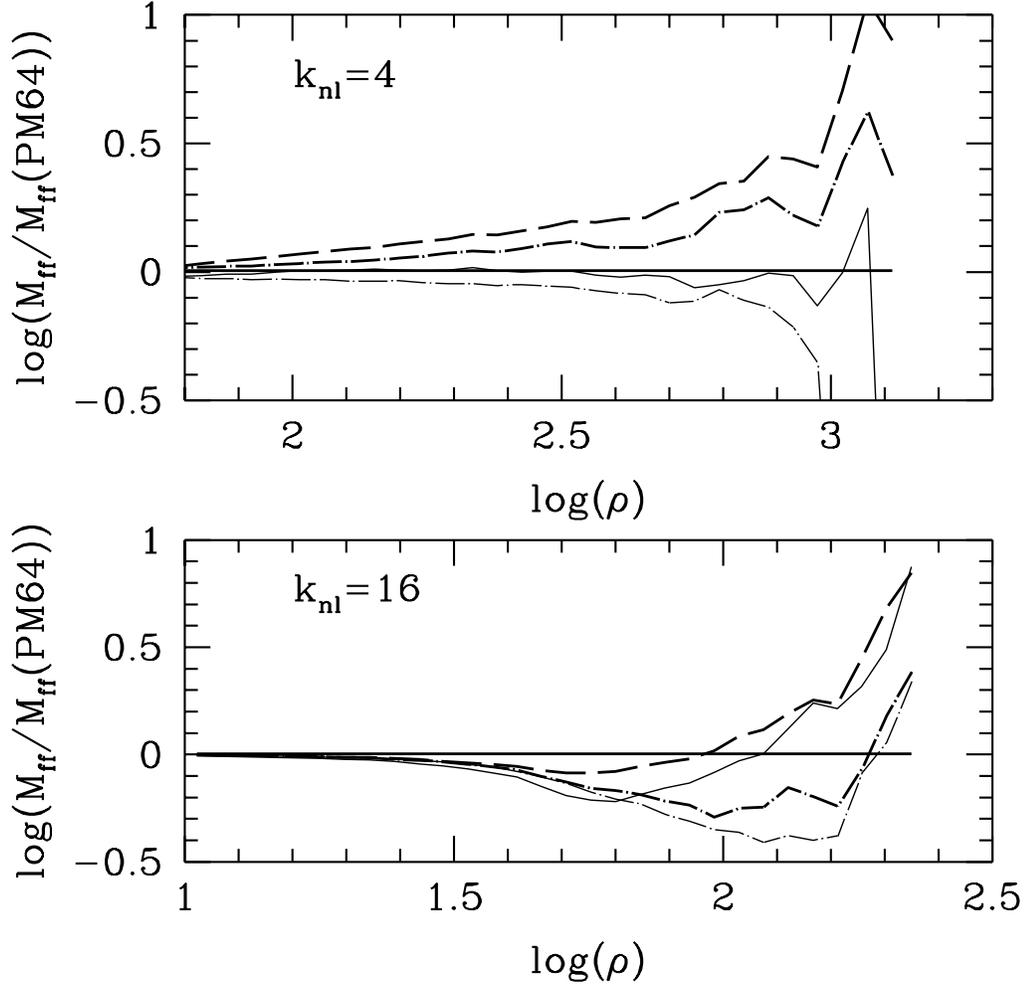,height=16cm}
\end{center}
\caption{
  The lines (same types as in Figure 1) show the amount of mass
  in regions of densities greater than the density threshold shown on
  the horizontal. The densities were calculated on the $128^3$ mesh.
  The values are shown as ratios to the fiducial
  PM model with $N=128^3$, and $ k_c = 64$, at our earliest evolved stage
(bottom)
 and our latest (top).
\label{fig:fig3}}
\end{figure}

\end{document}